# Quantum Algorithm for Searching of Two Sets Intersection

Kamil Khadiev[a,] *(http://orcid.org/0000-0002-5151-9908)[a], and Elizaveta Krendeleva[a]

[a]*Kazan Federal University, Institute of Computational Mathematics and Information Technologies, Kazan, Russia*
*e-mail: kamilhadi@gmail.com

**Abstract**—In the paper we investigate Two Sets Intersection problem. Assume that we have two sets that are subsets of n objects. Sets are presented by two predicates that show which of n objects belong to these sets. We present a quantum algorithm that finds an element from the two sets intersection. It is a modification of the well-known Grover's search algorithm that uses two Oracles with access to the predicates. The algorithm is faster than the naive application of Grover's search.

## 1. INTRODUCTION

Quantum computation and quantum algorithms [1] are one of the hot topics in the last decades. There are many problems where we can obtain a quantum speed-up. Some of them can be founded here [2,3]. One of the well-known quantum algorithms is the Grover's search algorithm for a Search Problem. The Grover's search algorithm [4,5] is a well-known algorithm for the Search Problem. It has several modifications and generalizations [4-11]. It has wide area of applications some of them are listed here [17-23]. Versions with possible errors in input data were also considered [12-14]. The standard setting for the algorithm is search an element in a search space of size n that satisfy some predicate $f_X$, i.e. we search an element i from 0 to n-1 such that $f_X(i)$=true. The problem can be interpreted as a searching an element from a set X where X is a subset of $\{0,…,n-1\}$. Here X=$\{i: f_X(i)$=true, $0 \leq i \leq n-1\}$. The Grover's search algorithm finds an element in $O((n/|X|)^{0.5})$ steps. Each step is a pair of two unitary matrices that are a query to the oracle $O_X$ and the diffusion D. Here diffusion is a rotation of all amplitudes near the mean; $O_X$ is access to an oracle that can compute $f_X(i)$.

In the paper, we consider the Two Sets Intersection problem with superset setting. We assume that our algorithm has access to two predicates $f_X$ and $f_Y$ that correspond to two sets X=$\{i: f_X(i)$=true, $0 \leq i \leq n-1\}$ and Y=$\{i: f_Y(i)$=true, $0 \leq i \leq n-1\}$ such that Y⊂X. Our goal is to find an element that belongs to both sets. Formally, we search i such that $f_X(i)$=true and $f_Y(i)$=true, in other words i is from intersection of X and Y.

We can solve the problem by applying the Grover's search algorithm for the predicate $f_Y$. We remember that Y⊂X, that is way any element from Y belongs to the intersection. In that case, we can find a target element in $O((N/|Y|)^{0.5})$ queries to $f_Y$ using Grover's search algorithm. Let $T_Y$ be complexity of computing $f_Y$. Then, the total complexity of this solution is $O((N/|Y|)^{0.5}T_Y)$.

Here, we assume that the complexity of computing $f_Y$ is much more than complexity of computing $f_X$, in other words, $T_Y \gg T_X$. So, it is important to minimize the number of queries to

$f_Y$. In the paper, we suggest an algorithm that does a constant number of queries to $f_Y$, and has $O((N/|X|)^{0.5}T_X+T_Y)$ complexity where $T_X$ is complexity of computing $f_X$. It is better than the naive usage of Grover's algorithm in the case of $T_Y \gg T_X$.

The structure of the paper is following. Section 2 contains prelimenaries. A description and analysis of an algorithm are in Section 3. Conclusion is presented in Section 4.

## 2. PRELIMINARIES

Let us present the formal definition of the problem.

**Problem.** For an integer n, we consider two functions $f_X:\{0,\ldots,n-1\}\to\{\text{false, true}\}$ and $f_Y:\{0,\ldots,n-1\}\to\{\text{false, true}\}$. We want to find any $i\in\{0,\ldots,n-1\}$ such that $f_X(i)=\text{true}$ and $f_Y(i)=\text{true}$. We solve the problem with assumption that for any j such that $f_Y(i)=\text{true}$, we have $f_X(i)=\text{true}$.

The set interpretation is the following. Let a set $X=\{i:i\in\{0,\ldots,n-1\} \text{ and } f_X(i)=\text{true}\}$, and $Y=\{i:i\in\{0,\ldots,n-1\} \text{ and } f_Y(i)=\text{true}\}$. In other words, the function $f_X$ is a characteristic function of the set X, and the function $f_Y$ is a characteristic function of the set Y. We want to find any $i\in X\cap Y$. We solve the problem with assumption that $Y\subset X$.

**Query model.** One of the most popular computation models for quantum algorithms is the query model. We use the standard form of the quantum query model. Let $g:D\to\{0,1\}$, $D\subset\{0,1\}^M$ be an M variable function. We wish to compute on an input $t\in D$. We are given oracle access to the input t, i.e. it is implemented by a specific unitary transformation that is usually defined as $|i\rangle |z\rangle |w\rangle \to |i\rangle |z+x_i \bmod 2\rangle |w\rangle$ where the $|i\rangle$ register indicates the index of the variable we are querying, $|z\rangle$ is the output register, and $|w\rangle$ is some auxiliary work-space. The operation is implemented by the CNOT gate. An algorithm in the query model consists of alternating applications of arbitrary unitaries independent of the input and the query unitary, and measurement in the end. The smallest number of queries for an algorithm that outputs g(x) with probability at least 2/3 on all x is called the quantum query complexity of the function f, and Q(f) notation is used for it.

We refer the readers to [15,16,7] for more details on quantum query model and quantum computing [1].

## 3. ALGORITHM

Let us present the algorithm for the problem. The algorithm is based on Grover's search algorithm [4,5] and uses elements of this algorithm. We assume that we have oracle access to the input and we can compute functions $f_X$ and $f_Y$.

We assume that the function $f_X$ can be computed and complexity is $T_X$. We use three registers $|\text{ind}\rangle|r\rangle|b_x\rangle$. The first register contains $\log_2 n$ qubits and has states from $|0\rangle$ to $|n-1\rangle$. The second register contains a single qubit. The third register is some auxiliary work-space for computing $f_X$.

We have a unitary $O_X$ that does the following action
$$O_X : |i\rangle|r\rangle|b_x\rangle \to |i\rangle|(r + f_X(i)) \bmod 2\rangle|b_x\rangle.$$

Before the algorithm we prepare the register $|a\rangle$ in the following form.
$$|r\rangle = (|0\rangle - |1\rangle)/2^{0.5}$$

We can do it by applying inversion X matrix and Hadamard H matrix to the initial $|0\rangle$ state. In that case, the unitary $O_X$ does the following action
$$O_X : |i\rangle|r\rangle|b_x\rangle \to (-1)^{f_X(i)} |i\rangle|r\rangle|b_x\rangle.$$
The complexity of applying $O_X$ is $T_X(n)$.

Similarly, we have a unitary $O_Y$ with complexity $T_Y(n)$ that does the following action
$$O_Y : |i\rangle|r\rangle|b_y\rangle \to (-1)^{f_Y(i)} |i\rangle|r\rangle|b_y\rangle.$$
Here $|b_y\rangle$ is some auxiliary work-space for computing $f_Y$.

One more unitary that is used in the algorithm is Grover's diffusion D. The unitary rotates all amplitudes near a mean.
$$D = H^{\otimes \log n}(2|0...0\rangle\langle 0...0| - I^{\otimes \log n})\ H^{\otimes \log n}$$
Here H is Hadamard matrix, $I^{\otimes \log n}$ is nxn identity matrix, $|0...0\rangle$ is the zero state for $\log_2 n$ qubits. We apply the transformation to $|ind\rangle$ register by the following way. If $|ind\rangle=d_0|0\rangle+...+d_{n-1}|n-1\rangle$ and $m=(d_0+...+d_{n-1})/n$, then
$$D: d_0|0\rangle+...+d_{n-1}|n-1\rangle \to (2m-d_0)|0\rangle+...+(2m-d_{n-1})|n-1\rangle$$

Let us fix some parameter L which value we discuss later. The algorithm has the initial phase and three main phases. It starts from the initial $|ind\rangle|r\rangle|b_x\rangle|b_y\rangle=|0...0\rangle|0\rangle|0...0\rangle|0...0\rangle$ state.

Phase 0. Firstly, we apply inversion X matrix to $|r\rangle$ register. Then, we apply Hadamard transformation to all qubits of $|ind\rangle|a\rangle$ registers. So, the state before the main phases is the following one
$$|ind\rangle=(1/n^{0.5})(|0\rangle+...+|n-1\rangle),\ |r\rangle=1/2^{0.5}(|0\rangle+|1\rangle).$$

Phase 1. One step of the first phase is applying $O_X$ and D transformations. On this phase we do L steps.

Phase 2. One step of the second phase is applying $O_Y$ and D transformations. On this phase we do the single step.

Phase 3. One step of the third phase is applying $O_X$ and D transformations. On this phase we do 2L steps.

After that we measure $|ind\rangle$ register and obtain the index i of target element such that $f_X(i)$=true and $f_Y(i)$=true.

*Complexity of the Algorithm*

Let us discuss query complexity of the algorithm. The analysis is motivated by analysis of the Grover's search algorithm. We focus on amplitudes of qubits in register $|ind\rangle$. Let it be
$$|ind\rangle=a_0|0\rangle+...+a_{n-1}|n-1\rangle.$$

Let us split all indexes from $\{0,…,n-1\}$ set to three groups:
- $K_{11}=\{i: 0\leq i\leq n-1$ and $f_X(i)$=true and $f_Y(i)$=true$\}$,
- $K_{10}=\{i: 0\leq i\leq n-1$ and $f_X(i)$=true and $f_Y(i)$=false$\}$,
- $K_{00}=\{i: 0\leq i\leq n-1$ and $f_X(i)$=false and $f_Y(i)$=false$\}$.

After Phase 0 all values $a_i$ are equal. Possible transformations on Phases 1-3 are $O_X$, $O_Y$ and D. It is easy to see that all values $a_i$ for $i\in K_{11}$ are the same on each step of the algorithm. Similar claim is correct for two other sets. So, let $a_{11}=a_i$ for $i\in K_{11}$, $a_{10}=a_i$ for $i\in K_{10}$, and $a_{00}=a_i$ for $i\in K_{00}$.

We know that on each step $a_0^2+...+a_{n-1}^2=1$. So, we have
$$(a_{11}^2+...+a_{11}^2) + (a_{10}^2+...+a_{10}^2) + (a_{00}^2+...+a_{00}^2) = 1$$

$$|K_{11}|a_{11}^2 + |K_{10}|a_{10}^2 + |K_{00}|a_{00}^2 = 1$$
$$(|K_{11}|^{0.5}a_{11})^2 + (|K_{10}|^{0.5}a_{10})^2 + (|K_{00}|^{0.5}a_{00})^2 = 1$$

Let $z = |K_{11}|^{0.5}a_{11}$, $y = |K_{10}|^{0.5}a_{10}$, and $x = |K_{00}|^{0.5}a_{00}$. So, we have $z^2 + y^2 + x^2 = 1$ on each step. We can say, that on each step the state of the quantum register can be associated with a point $(x,y,z)$ in three dimensional space. The point is always on a unit sphere.

Let us look to the $O_X$, $O_Y$ and $D$ and how they affect the $(x,y,z)$ point on the unit sphere. The transformation $O_X$ is such that

$$O_X : a_i \rightarrow -a_i \text{ for } i \in K_{11} \cup K_{10}.$$

Therefore, the transformation affects $(x,y,z)$ in the following way:

$$O_X : (x,y,z) \rightarrow (x,-y,-z).$$

We can say, that this transformation is a reflection of the point relative to the X-axis.

The transformation $O_Y$ is such that

$$O_Y : a_i \rightarrow -a_i \text{ for } i \in K_{11}.$$

Therefore, the transformation affects $(x,y,z)$ in the following way:

$$O_X : (x,y,z) \rightarrow (x,y,-z).$$

We can say, that this transformation is a reflection of the point relative to the XY-plain.

According to the analysis from [4,5,7], we know that the transformation D is a reflection of the vector $|ind\rangle$ relative to the vector $|P\rangle = (1/n^{0.5})(|0\rangle + ... + |n-1\rangle)$. In our terms, the transformation reflects the point relative to the point $S(x_s, y_s, z_s)$, where

$$z_s = (|K_{11}|/n)^{0.5}, \quad y_s = (|K_{10}|/n)^{0.5}, \text{ and } x_s = (|K_{00}|/n)^{0.5}.$$

Our goal is maximization of sum of squares of amplitudes for states with indexes $i \in K_{11}$ after Phase 3. If we do it, then after measurement in the last step, we obtain one of the states $|i\rangle$ such that $f_X(i) = \text{true}$ and $f_Y(i) = \text{true}$. We can say, that our goal is maximization of z coordinate for our point on the unit sphere.

The following analysis is based on computational experiments. Let us look to the trace of the point on the sphere (Figure 1).

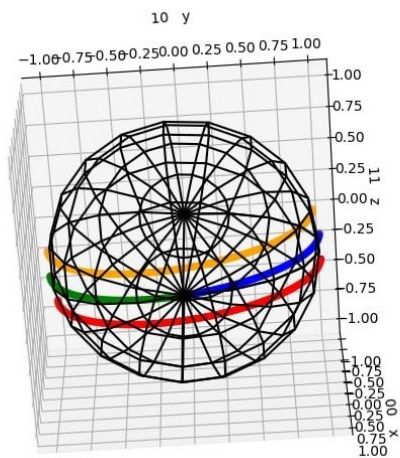

Figure 1. Trace of the point.

Here blue and green lines for the Phase 1; yellow and red lines for the Phase 2. The blue line is points after a step that is $O_X$ and $D$ on Phase 1. The green line is points after only one part of a step that is $O_X$. The yellow line is points after a step that is $O_X$ and D on Phase 2. The red line is points after only one part of a step that is $O_X$.

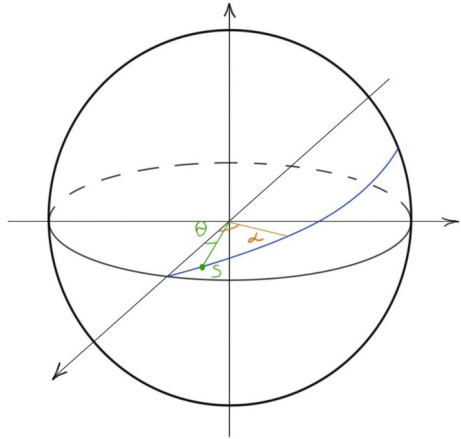

Figure 2. Trace of the point on Phase 1.

Due to picture analysis and computational experiments, we can see that the best result can be achieved when α is close to π/2 on Figure 2 where we present positions of the point on Phase 1. Let us compute L such that the claim is true.

If we consider the circle that contains the blue line, then similar to the analysis for the Grover's search algorithm [4,5,7], we can say, that the $O_X$ transformation flips the sign of the angle α, i.e. $O_X$: α → - α. The transformation D is such that α → 2θ - α. So, after a step we obtain α → α+2θ. In L steps we obtain the angle α=(2L+1)θ. We want to obtain α≈π/2, therefore, we can get L≈(π/4)/ θ. Let us look to the angle θ closely on Figure 3.

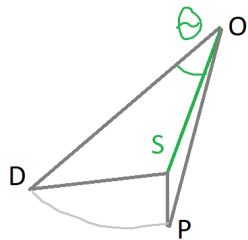

Figure 3. The point S after applying Hadamard transformation and the corresponding angle θ.

Because of the assumption $|Y|<<|X|$, the angle θ is very small. Therefore, we can say that DP is approximately a line. In that case, DO=1, SO=1, SP=$z_s$=$(|K_{11}|/n)^{0.5}$, DP= $y_s$= $(|K_{10}|/n)^{0.5}$, DSP is an orthogonal triangle; DOS is an isosceles triangle. So we can compute DS by the Pythagorean theorem:

$$DS=(SP^2 + DP^2)^{0.5}=((|K_{11}|+|K_{10}|)/n)^{0.5}.$$

So we can compute θ by the following formula:

$$θ=2\arcsin(0.5DS/DO)=2\arcsin(0.5DS)=2\arcsin(0.5((|K_{11}|+|K_{10}|)/n)^{0.5})$$

We remember that θ is very small. So, due to the first remarkable limit, we can say that 0.5 θ ≈arcsin (0.5 θ).Therefore, we have

$$θ= 2 \cdot 0.5((|K_{11}|+|K_{10}|)/n)^{0.5}=((|K_{11}|+|K_{10}|)/n)^{0.5}=(|X|/n)^{0.5}.$$

Finally, we can compute L by the following formula:

$$L≈(π/4)/ θ= L≈(π/4)(n/|X|)^{0.5}=O((n/|X|)^{0.5}).$$

We can say that we have $O((n/|X|)^{0.5})$ queries to $f_X$ and a single query to $f_Y$. In fact, we prove the following theorem.

**Theorem 1.** *The presented algorithm finds an element from the set Y with $O((n/|X|)^{0.5}T_X+T_Y)$ query complexity.*

According to computational experiments, the success probability can be increased and reach probability close to 1 if we repeat the algorithm 10-20 times.

## 5. CONCLUSION

We presented an algorithm for the Two Sets Intersection problem that has $O((n/|X|)^{0.5}T_X+T_Y)$ in the case of $Y \subset X$ and $|Y|<<|X|$. That is better than the simple application of Grover's search algorithm if $T_X>>T_Y$.

The open problems are developing algorithm for the case $Y/X \neq \emptyset$; and removing the restriction $|Y|<<|X|$.

## FUNDING

This research was funded by the subsidy allocated to Kazan Federal University for the state assignment in the sphere of scientific activities, project No. 0671-2020-0065.

## CONFLICT OF INTEREST

The authors declare that they have no conflicts of interest.

## REFERENCES

1. Nielsen, Michael A., and Isaac L. Chuang. *Quantum computation and quantum information*. (Cambridge university press, 2010).

2. De Wolf, Ronald. *Quantum computing and communication complexity.* (University of Amsterdam, 2001).

3. Jordan, Stephen. *Quantum Algorithms Zoo.* http://quantumalgorithmzoo.org/, access on 10 November 2023).

4. Brassard, Gilles, Peter Hoyer, Michele Mosca, and Alain Tapp "Quantum amplitude amplification and estimation." Contemporary Mathematics **305**, 53-74 (2002).

5. Grover, Lov K. "A fast quantum mechanical algorithm for database search." In *Proceedings of the twenty-eighth annual ACM symposium on Theory of computing*, (1996), pp. 212-219.

6. Long, G. L. . Grover algorithm with zero theoretical failure rate. Physical Review A, 64(2), 022307 (2001).

7. Khadiev, Kamil. "Lecture notes on quantum algorithms." arXiv preprint arXiv:2212.14205 (2022).

8. Kothari, Robin. "An optimal quantum algorithm for the oracle identification problem." In 31st International Symposium on Theoretical Aspects of Computer Science, p. 482. 2014.

9. Lin, C. Y. Y., & Lin, H. H. " Upper Bounds on Quantum Query Complexity Inspired by the Elitzur--Vaidman Bomb Tester." Theory of Computing, 12(1), 1-35 (2016).

10. Dürr, C., Heiligman, M., HOyer, P., & Mhalla, M. "Quantum query complexity of some graph problems." SIAM Journal on Computing, 35(6), 1310-1328 (2006).

11. Kapralov, R., Khadiev, K., Mokut, J., Yagafarov, M., & Shen, Y. "Fast Classical and Quantum Algorithms for Online k-server Problem on Trees". In CEUR Workshop Proceeding, 3072, pp. 287-301 (2021).

12. Høyer, P., Mosca, M., & De Wolf, R. "Quantum search on bounded-error inputs." In *International Colloquium on Automata, Languages, and Programming*, pp. 291-299 (2003).

13. Ambainis, A., Bačkurs, A., Nahimovs, N., & Rivosh, A. " Grover's algorithm with errors." In *International Doctoral Workshop on Mathematical and Engineering Methods in Computer Science*, pp. 180-189 (2012).

14. Kravchenko, D., Nahimovs, N., & Rivosh, A. " Grover's search with faults on some marked elements.", International Journal of Foundations of Computer Science, **29**(04), 647-662 (2018).

15. Ambainis, A. "Understanding quantum algorithms via query complexity." In *Proceedings of the International Congress of Mathematicians: Rio de Janeiro 2018*, pp. 3265-3285 (2018).

16. Ablayev, F., Ablayev, M., Huang, J. Z., Khadiev, K., Salikhova, N., & Wu, D. "On quantum methods for machine learning problems part I: Quantum tools."Big data mining and analytics, 3(1), 41-55.

17. Khadiev, K., Ilikaev, A., & Vihrovs, J. "Quantum algorithms for some strings problems based on quantum string comparator. " Mathematics, **10**(3), 377 (2022).

18. Aborot, J. A. "An Oracle Design for Grover's Quantum Search Algorithm for Solving the Exact String Matching Problem." In *Theory and Practice of Computation: Proceedings of Workshop on Computation: Theory and Practice WCTP2017,* pp. 36-48 (2019).

19. Khadiev, K., & Remidovskii, V. " Classical and quantum algorithms for constructing text from dictionary problem." Natural Computing, **20**(4), 713-724 (2021).

20. Le Gall, F., & Seddighin, S. "Quantum meets fine-grained complexity: Sublinear time quantum algorithms for string problems".Algorithmica, 85(5), 1251-1286. (2023)

21. Le Gall, F., & Seddighin, S. " Quantum Meets Fine-Grained Complexity: Sublinear Time Quantum Algorithms for String Problems. "In *13th Innovations in Theoretical Computer Science Conference (ITCS 2022).* Schloss Dagstuhl-Leibniz-Zentrum für Informatik. (2022).

22. Ramesh, H., & Vinay, V. "String matching in O ($\sqrt{n}$+ $\sqrt{m}$) quantum time." Journal of Discrete Algorithms, **1**(1), 103-110 (2003).

23. Khadiev, K., & Safina, L. "Quantum Algorithm for Dynamic Programming Approach for DAGs and Applications." Lobachevskii Journal of Mathematics, 44(2), 699-712 (2023).